\documentclass[showpacs,superscriptaddress,twocolumn,aps,amsmath,amssymb]{revtex4}

\usepackage{amsfonts}
\usepackage{amsmath}
\usepackage{graphicx}
\usepackage{dcolumn}
\usepackage{bm}

\newcommand{\be}{\begin{equation}}
\newcommand{\ee}{\end{equation}}
\newcommand{\bea}{\begin{eqnarray}}
\newcommand{\eea}{\end{eqnarray}}
\newcommand{\Fig}[1]{Fig.\ref{#1}}
\newcommand{\Eq}[1]{Eq.\,(\ref{#1})}

\newcommand{\dg}{\dagger}

\newcommand{\nl}{\nonumber \\}

\begin{document}

\title{Bias-induced insulator-metal transition in organic electronics}
\author{J.~H.~Wei}
\affiliation{Department of Physics, Renmin University of China,
Beijing, China} \affiliation{Department of Physics, Shandong
University, Jinan, China}

\author{S.~J.~Xie}
\affiliation{Department of Physics, Shandong University, Jinan,
China}

\author{L.~M.~Mei}
\affiliation{Department of Physics, Shandong University, Jinan,
China}

\author{YiJing Yan}
\affiliation{Department of Chemistry, Hong Kong University
   of Science and Technology, Kowloon, Hong Kong}

\date{\today}

\begin{abstract}
 We investigate the bias-induced insulator-metal transition
 in organic electronics devices, on the basis of the
 Su-Schrieffer-Heeger model [W.~P. Su, J.~R. Schrieffer, and A.~J. Heeger,
 Phys. Rev. B {\bf 22}, 2099 (1980)] combined with the non-equilibrium
 Green's function formalism. The insulator-metal
 transition is explained with the energy levels
  crossover that eliminates the Peierls phase [R.~Peierls,
{\em Quantum Theory of Solids}, Oxford University Press, Oxford,
1955] and
 delocalizes the electron states near the
 threshold voltage. This may account for
 the experimental observations on the devices
 that exhibit intrinsic bistable conductance switching
 with large on-off ratio.
\end{abstract}


\maketitle
\clearpage

  In recent years molecular electronics,
 which uses individual molecules as possible electrical switches,
 has grown up rapidly since the discoveries of negative
 differential resistance and
 bistable conductance switching (CS) \cite{Che991550,Don012303}.
 Individual molecules can be wired to serve as diodes, transistors,
 and other electronic devices at the heart of
 computer chips \cite{Ree013735,Col001172,Kra022927,Raw03377}.
 For the next generation of technology,
 molecular electronics should exhibit
 a large on-off ratio ($>$\,50:1).
 It should also be intrinsic and controllable, rather than
 artificial due to for example metal filament \cite{Ser03556}.
  Some mechanisms such as the conformational change \cite{Don012303},
 charging effect and polaron models \cite{Gal05125,Wei0682} have
 been proposed. However, none of them can fully account
 for the large intrinsic on-off ratio
 of molecular switching observed
 in experiments \cite{Blu05167,Kum991645,Mat0512450}.
 In this work, we shall propose another possible
 mechanism, in which the
 intrinsic CS with large on-off ratio
 is controlled by the bias-triggered
 Peierls phase elimination and energy levels crossover.

  Let us start with the following observations.
 Most molecular devices use short-chain $\pi$-conjugated
 organic molecules as the active region, and are
 fabricated as metal-molecule-metal sandwiched structures.
 A chain-like nanoscale organic molecule belongs to
 the family of organic semiconductors (OSEs).
 Physically, an OSE is charactered by the Peierls instability;
 i.e., the strong electron-phonon ({\it e-ph})
 interaction that induces the single--double bond alternation
 and a relative large energy gap between highest-occupied and
 lowest-unoccupied molecular orbital (HOMO and LUMO) \cite{Pei55}.
 It is the Peierls instability that largely localizes
 the $\pi$-electrons of carbon and contributes to
 the low conductance of OSEs, no matter
 they are fabricated via the bottom-up self-assembled monolayer
 or the top-down lithography technique.
 Apparently, delocalizing $\pi$-electrons will improve
 the conductivity of OSEs. One method is the creation of
 nonlinear excitations (solitons or polarons) to act as charge
 carriers \cite{Su802099}.  Another approach
 is to eliminate the Peierls phase directly.
  Increasing temperature may
  melt the Peierls phase, leading to an insulator--metal
 (I-M) transition \cite{Pei55}.

 An effective and controllable method
 in electronics should however be the one triggered
 by the bias or other external control field.
%%%
 In this connection and also in contact
 with some typical
 experimental systems \cite{Kum991645,Mat0512450},
 we present in this work a theoretical study
 on the bias-induced I-M transition in a model
 nanoscale metal/OSE/metal sandwich structure.
 We will elucidate the fact that the bias-triggered
 energy levels crossover and its resulting Peierls phase
 elimination can largely account for the observed CS.

 We combine the celebrated Su-Schrieffer-Heeger (SSH) model \cite{Su802099}
 and the nonequilibrium Green's function (NEGF)
 formalism \cite{Kel641515,Bra02165401} to study organic
 electronics. For the spin-independent charge transport considered
 in this work, the SSH Hamiltonian
 for the OSE electrons coupled adiabatically with the lattice
 displacements assumes
\bea\label{SSH}
 H_{O}&=&\sum_{n}
  \Bigl\{\epsilon _{o}c_{n}^{+}c_{n}-
  [t_{o}-(-1)^{n}t_{1}-\alpha _{o}y_{n}]\times
\nl &\ &
  (c_{n}^{+}c_{n+1}+
  c_{n+1}^{+}c_{n})\Bigr\}+\frac{K_{o}}{2}\sum_{n}y_{n}^{2}.
\eea%
 Here, $c_{n}^{+}$ ($c_{n}$) denotes the creation
(annihilation) operator of an electron at the $n^{\mathrm{th}}$
 site of the OSE, and  $\epsilon _{o}$ the on-site energy,
 $t_{o}$ the zero-displacement hopping integral,
 and $t_{1}$ the nondegeneracy parameter,
 respectively. The lattice distortion is written
 in terms of the bond distances $\{y_{n}=u_{n+1}-u_{n}\}$
 and treated classically, with
 the spring constant $K_{o}$
 and the adiabatic {\it e-ph}\
 coupling constant $\alpha _{o}$.
 The L and R electrodes are chosen to be an identical nonmagnetic metal.
 They are individually described
 by the one-dimensional single-band tight-binding Hamiltonian,
 being of the on-site energy $\epsilon _{f}$ and the
 nearest neighbor transfer integral $t_{f}$.

 At the mean-field level of correlations, one can evaluate the
 current based on the NEGF formalism \cite{Kel641515,Bra02165401},
\begin{equation} \label{I-V}
  I=\frac{2e}{h}\int\limits_{-\infty }^{\infty }
  {\mbox{Tr}\left( \Gamma _{\mathrm{L}}G^{\rm r}
    \Gamma _{\mathrm{R}}G^{\rm a}\right) }
   [f(E,\mu _{\mathrm{L}})-f(E,\mu _{\mathrm{R}})]dE.
\end{equation}%
 Here, $f(E,\mu _{\mathrm{L/R}})$ is the Fermi distribution
 function at the lead chemical potential $\mu _{\mathrm{L/R}}$.
 The trace term in the integrand is the transmission coefficient
 function $T(E)$. The involving
 $\Gamma _{\mathrm{L/R}}$ denotes the lead reservoir-induced
  broadening matrix, while
  $G^{\rm r}$ and  $G^{\rm a}=(G^{\rm r})^{\dg}$ are the retarded
 and the advanced single-particle Green's functions
 for the central scattering region (S-region),
 under a finite applied bias. This region
 consists of the OSE together with a number of
 metal atoms attached to each of its ends.
%%%
   For a quantum open system, the effects
 of both the charging from the electrodes and the
 external potential from the bias voltage
 are contained in the
 nonequilibrium reduced density matrix,
 $\rho =\frac{1}{2\pi }\sum_{\alpha
  =\mathrm{L,R}}  %\int\limits_{-\infty }^{\infty }
   \int G^{\rm r}\Gamma _{\alpha}G^{\rm a}
   f(E,\mu _{\alpha})dE$,
 of the coupled {\it e-ph}\ system \cite{Bra02165401}.
 The reduced density matrix should be evaluated
 in a self-consistent manner,
  together with the Poisson's
 equation and the Hellman-Feynman (H-F)
 variation theorem for lattice distortion:
 $\partial \lbrack \mathrm{Tr}%
 (H_{\mathrm{S}}\rho )]/\partial u_{n}=0$. The self-consistent
 procedure as well as the time-to-voltage mapping method to handle
 the time-dependent bias sweeping have been described
 in detail elsewhere \cite{Wei0682,Wei0508417}.

   In the following calculation of the hysteretic $I-V$ characteristics,
we choose a constant rate of bias sweeping to be $0.1$ V/sec, as
both the numerical efficiency and the typical experimental value
being of $0.01\sim 0.1$ V/sec are concerned. Note also that
SSH+NEGF formulation includes two main approximations. The first
one is the static lattice description involving in the model
Hamiltonian and the aforementioned H-F variation theorem. This
approximation is valid when the lattice fluctuation time scale is
longer than the inverse coherent charge transfer coupling. The
second approximation is the meanfield treatment of
electron-electron correlation that involves in the NEGF
formulation and also the model Hamiltonian. This is a common
approach, especially in treating strong
 {\it e-ph} systems.

 We choose the SSH parameters in \Eq{SSH} the
typical values:
 $t_{o}=2.5$ eV, $t_{1}=0.04$ eV, $\alpha _{o}=4.2$ eV/\r{A},
 and $K_{o}=21.0$ eV/\r{A}$^{2}$.
 The transfer integral for metal electrodes
 is $t_{f}=3.0$ eV to produce a wide
 band (4$t_{f}$), and the metal-OSE
 coupling transfer integral is
 $t_{\rm metal-OSE}=0.5(t_{o}+t_{f})$.
 Without losing generality,
 we set  $E_{\rm F}=\epsilon _{o}=\epsilon_{f}=0$,
 where $E_{\rm F}$ is the equilibrium Fermi level at
 the zero bias $V=0$.

\begin{figure}
 \includegraphics[width=3.in]{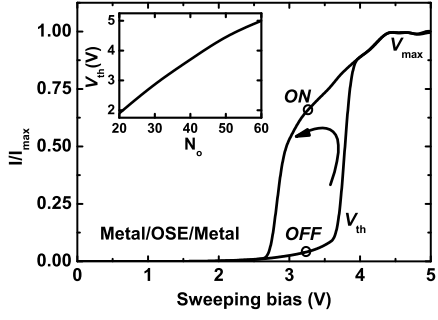}
 \caption{The hysteretic current as the function of sweeping bias
 of the model metal/OSE/metal electronics at $T=11$ K.
 The size of OSE is $N_o=40$ and the linear voltage
 sweep rate is at 0.1 V/sec.
 The insert is the threshold voltage $V_{\rm th}$
 as a function of the OSE chain
 length $N_o$.}
 \label{fig1}
 \end{figure}

 We then calculate the nonlinear current [\Eq{I-V}] as a function
 of sweeping bias of the model metal/OSE/metal electronics at
 $T=11$ K by using the time-to-voltage mapping method \cite{Wei0508417}.
 The resulting $I$-$V$ curve, presented in
\Fig{fig1}, clearly shows hysteretic CS characters.
 Note that the model OSE is of $N_o=40$, which is about 5nm
 and comparable to the
 experiments in Refs.\ \onlinecite{Blu05167} and
 \onlinecite{He0663}.
 More than 30 years ago, Simmons
 \emph{et al.}\ reported an
 electroformed metal-insulator-metal diode
 with reversible voltage memory effects \cite{Sim6777}.
 Recently, Bozano {\it et al.}\
 had also demonstrated a similar resistive
 switching phenomenon in OSE layers \cite{Boz04607}.
 They had proposed several necessary features of an operational
 switch, as summarized in Ref.\ \onlinecite{Boz04607}.
 In comparison with their experiments,
 especially the Fig.\,3
 in Ref.\ \onlinecite{Boz04607},
 our model calculation dose possess most of the necessary features
 they proposed. These include
 (1) a distinct bistable on/off state
 corresponding to low/high resistance with
  an on-off ratio $\gtrsim$ 20:1;
 (2) a local maximum in current $I_{\rm max}$ which
 nonlinearly depends on the size of OSE and the metal-OSE
 coupling (varying from several nA $\sim{\rm \mu}$A) at voltage $V_{\rm
 max}$ ($\sim$ 4.5 V);
 (3) a threshold of voltage $V_{\rm th}$
 ($\sim$ 3.7 V) at which the high-current
 (switch-on) state is established;
 (4) reproducible switching and reading which can be
 achieved with applied positive or negative voltage.
  The insert in \Fig{fig1} shows the
  nonlinear dependence of the threshold voltage
  $V_{\rm th}$ on the OSE chain length $N_o$.
 This confirms the generality of our results
 in organic electronics devices with different sizes.

\begin{figure}
 \includegraphics[width=3. in]{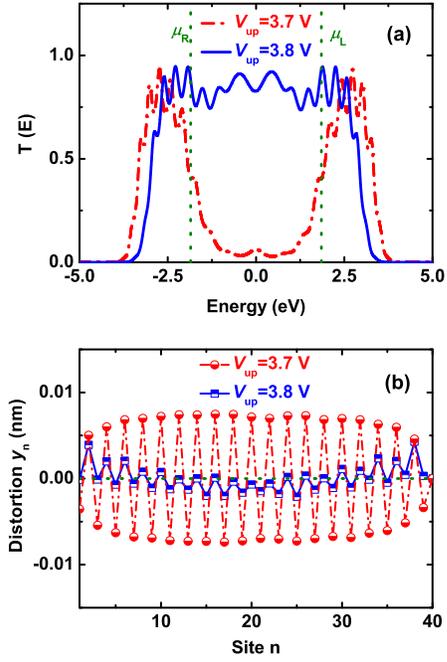}
 \caption{(color online) High (solid lines) and low (dash-dot lines) conductance
 states triggered  by the up-sweeping bias.
 (a) Transmission coefficient $T(E)$. The vertical
 dot lines indicate the window between
 $\mu_{\rm L}$ and $\mu_{\rm R}$ for the charge
 transmission; (b) Lattice distortion $y_n\equiv u_{n+1}-u_n$
 of the OSE substructure.}
 \label{fig2}
 \end{figure}

\begin{figure}
 \includegraphics[width=3. in]{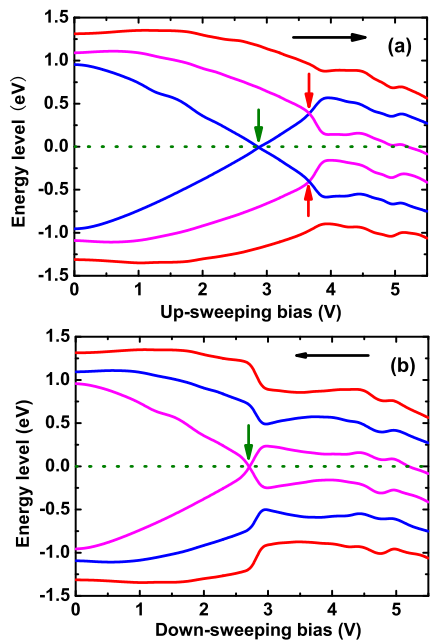}
 \caption{(color online) The voltage-evolution of six energy levels
 of the OSE (HOMO-2, $\cdots$, LUMO+2) for up-sweeping (a) and
 down-sweeping bias (b). The arrows indicate the crossover of
 energy levels.}
 \label{fig3}
 \end{figure}

 Let us discuss the mechanism of the bistable CS,
 as indicated in \Fig{fig1}. The OSE substructure is found charge neutral,
 indicating there are no charging and polaron effects
 below the saturation voltage.
 The abrupt increase of current near the threshold
 $V_{\rm th}$ at 3.7 V should therefore be interpreted with
 an alternative mechanism.
  First of all, we shall show that the observed
 abrupt current change relates to an I-M transition.
 Presented in \Fig{fig2} are the changes of
 transmission coefficient $T(E)$ and lattice distortion $y_n$
 around the threshold voltage.
 Evidently, the Peierls phase of the OSE substructure
 is largely eliminated near $V_{\rm th}$.
 The key features in \Fig{fig2} include
 (1) the rapid increasing of transmission coefficient in
 the window between the indicated $\mu_{\rm L}$
 and $\mu_{\rm R}$ in \Fig{fig2}(a);
 (2) the distinct reducing of the lattice distortion
 or single-double bond alternation [\Fig{fig2}(b)];
 and also (3) the vanishing of the energy gap
 and the merging of valence and conductance bands,
 which will be explained later, cf.\ \Fig{fig3}(a).
 The above features all indicate an I-M transition
 around $V_{\rm th}$.
 When the bias sweeps down, the Peierls
 phase, as indicated by the lattice distortion
 in \Fig{fig2}(b) (see also \Fig{fig3} for energy gap),
 is reformed resulting in a metal-insulator (M-I)
 transition around 2.8 V.
 The delayed M-I transition, in comparing with
 the I-M transition voltage, is due to
 the lattice distortion reorganization,
 resulting in the hysteretic loop ($\sim 1$ V)
 in the $I$-$V$ curve in \Fig{fig1}.

  To further elucidate the bias-induced
 I-M/M-I transition mechanism, we also calculate the
 `reduced' energy levels and wave functions of the OSE substructure
 by diagonalizing its Hamiltonian. It consists of \Eq{SSH}, with
 the nonequilibrium lattice parameters being
 evaluated in the self-consistent
 manner as described earlier.
 The Hamiltonian contains also the
 electric potential variation in the OSE structure,
 which had also been determined self-consistently.
 The informations on this Hamiltonian are
 shown to be able to reflect the bias-induced intrinsic
 changes of the OSE in the device.

 The resulting energy levels from the
 aforementioned self-consistent Hamiltonian, as they vary
 with the up-sweeping and down-sweeping bias voltage,
 are reported in \Fig{fig3}, in which only
 six of them, HOMO-2, $\cdots$, LUMO+2, are examined, since they
 dominate most of the properties of the OSE during bias sweeping.
 The most striking feature of \Fig{fig3} is the bias-induced energy
 levels crossover. It occurs twice in the bias sweep-up
 [\Fig{fig3}(a)] but only once in the sweep-down direction
 [\Fig{fig3}(b)]. The first crossover in \Fig{fig3}(a) does not
 increase the current. Detailed analysis finds it due to the fact
 that the corresponding HOMO and LUMO states are localized near the
 metal-OSE interfaces by the surface vibration modes (not shown here
 due to space limitation). Note that the HOMO/LUMO here is a reduced state
 in the presence of electrodes and lattice.
 Unlike the first crossover, the second one, shown by the
 dual-crossing between the HOMO and LUMO+1
 and between the LUMO and HOMO-1 as indicated
 by the arrows in \Fig{fig3}(a) around $V_{\rm th}$,
 does increase the current, as it affects the system remarkably.
 It combines the
 valence and conductance bands of the OSE to form a single metallic
 band structure without energy gap;
 and meanwhile it also {\it completely
 delocalizes} the wave functions
 of those six levels over the whole OSE substructure
 The gap-free metallic band structure
 and the delocalized wave functions,
 together with the uniform valence bonds [cf.\ \Fig{fig2}(b)]
 around the second crossover,
 are highly advantageous to the transport of $\pi$-electrons.
 They lead to the I-M transition of the organic electronics
 with rapidly increasing current around the $V_{\rm th}$.

 In a similar but reversible manner,
 when the bias sweeps down the energy levels
 crossover at $V \sim 2.8$\,V
 re-opens the band gap [\Fig{fig3}(b)],
 re-localizes the electronic
 wave functions, and re-dimerizes the lattice displacements of the OSE.
 As results, the OSE substructure transforms
 from the metallic back to the insulating state (M-I transition)
 with the current sudden drop following
 a near-linear decrease (cf.\ \Fig{fig1}).

  Some experimental evidences of the bias-induced I-M transition
 have been reported in large size organic electronics.
 One is the bias-driven high-to-low resistive state
 transition in the
 alkali-7,7,8,8-tetracyanoquinodimethane (K-TCNQ) single crystals,
 a quasi-one dimensional organic charge-transfer complex
 (Mott-Peierls insulator) \cite{Kum991645}.
 The I-M transition occurs in this system
 at about several hundred of volt in low temperature, where
 a metallic path has been visualized with a microscope.
 The present study may also shed some light on understanding
  the experimental observations in Ref.\ \onlinecite{Mat0512450}.
 These include the conductance switching, hysteretic loop,
 and `dimerization' reduction in the TTF-based dimeric
  donor salt; cf.\ \Fig{fig1} here versus
  the Fig.~2 of Ref.\ \onlinecite{Mat0512450}.
  The detailed comparison between the
  theory and experiments will be published
 elsewhere.

 In summary, we have investigated the
 bias-induced I-M transition
 in organic electronics devices and
 explained it with the energy
 levels crossover which diminishes the Peierls phase and
 delocalizes the electron states
 at the threshold voltage.
 The bias-triggered intrinsic large on-off
 ratio in this work ($\gtrsim$ 20:1)
 is experimental controllable. Our theory study reproduces a
 range of experimental observations, such as
 the I-M phase transition \cite{Kum991645,Mat0512450},
 the hysteretic conductance switching \cite{Sim6777,Boz04607}.

Support from the National Natural Science Foundation of China
(Grants No.~10604037 and 10474056) and the Research Grants Council
of the Hong Kong Government (605105) is gratefully acknowledged.

\clearpage

\end{document}